\documentclass[12pt]{article}
\usepackage{openwork}
\usepackage{hyperref}
\usepackage{multicol}

\begin{document}

\title{Towards Sample-Efficient Entanglement Classification for 3–4 Qubit Systems: A Tailored CNN-BiLSTM Approach}

\author[1,2]{Qian Sun\textsuperscript{$\dagger$,}}
\author[3]{Yuedong Sun\textsuperscript{$\dagger$,}}
\author[1,2]{Yu Hu}
\author[1,2]{Yihan Ma}
\author[4]{Runqi Han}
\author[1,2]{Nan Jiang\textsuperscript{*,}}

\affil[1]{School of Physics and Astronomy, Beijing Normal University, Beijing 100875, China}
\affil[2]{Key Laboratory of Multiscale Spin Physics (Ministry of Education), Beijing Normal University, Beijing, 100875, China}
\affil[3]{Center for Quantum Information, IIIS, Tsinghua University, Beijing, 100084, China}
\affil[4]{School of Artificial Intelligence, China University of Petroleum (Beijing), Beijing 102249, China}

\affil[$^{\dagger}$]{These authors contributed equally to this work.}
\affil[$^*$]{jiangnan2019@bnu.edu.cn}

\date{}

\maketitle

\begin{abstract}
Accurate classification of multipartite entanglement in high-dimensional quantum systems is crucial for advancing quantum communication and information processing. However, conventional methods are resource-intensive, and even many machine-learning-based approaches necessitate large training datasets, creating a significant experimental bottleneck for data acquisition. To address this challenge, we propose a hybrid neural network architecture integrating Convolutional and Bidirectional Long Short-Term Memory networks (CNN-BiLSTM). This design leverages CNNs for local feature extraction and BiLSTMs for sequential dependency modeling, enabling robust feature learning from minimal training data. We investigate two fusion paradigms: Architecture 1 (flattening-based) and Architecture 2 (dimensionality-transforming).  When trained on only 100 samples, Architecture 2 maintains classification accuracies exceeding 90\% for both 3-qubit and 4-qubit systems, demonstrating rapid loss convergence within tens of epochs. Under full-data conditions (400\,000 samples), both architectures achieve accuracies above 99.97\%. Comparative benchmarks reveal that our CNN-BiLSTM models, especially Architecture 2, consistently outperform standalone CNNs, BiLSTMs, and MLPs in low-data regimes, albeit with increased training time. These results demonstrates that the tailored CNN-BiLSTM fusion significantly alleviates experimental data acquisition burden, offering a practical pathway toward scalable entanglement verification in complex quantum systems. \\

\end{abstract}

\section{Introduction}

The development of long-distance quantum communication networks, a central goal in quantum information science\cite{kimble_quantum_2008}, fundamentally relies on the reliable generation and verification of multipartite entanglement\cite{cirac_quantum_1997}. As envisioned in quantum repeater protocols\cite{duan_long_2001,briegel_quantum_1998,sangouard_quantum_2011}, scaling such networks requires distributing entanglement efficiently across many nodes\cite{bennett_teleporting_1993,bennett_purification_1996}, making the accurate classification of entanglement in high-dimensional systems a task of fundamental importance. However, this classification task faces a severe scaling challenge, due to the exponential scaling of required measurements\cite{guhne_entanglement_2009,cramer_efficient_2010,zhao_entanglement_2019,vedral_quantifying_1997,terhal_detecting_2002}. The resources required for traditional characterization\cite{bell_einstein_1964,clauser_proposed_1969}, such as those based on entanglement witnesses\cite{lewenstein_optimization_2000,sanpera_schmidt_2001} or positive partial transpose criteria\cite{peres_separability_1996,horodecki_separability_1996}, typically grow exponentially with the system size.\\

Machine learning (ML), particularly deep learning, has emerged as a powerful alternative and has demonstrated notable success across various quantum information tasks, including quantum state tomography\cite{glasser_neural_2018,torlai_neural_2018,an_unified_2024,urena_entanglement_2024,sun_enhanced_2025}, quantum circuits optimization\cite{fosel_quantum_2021,cao_quantum_2022}, and quantum control\cite{bukov_reinforcement_2018,an_deep_2019}. It also includes using the back-propagation algorithm\cite{samuel_studies_1959,lecun_gradient_1998,elman_finding_1990} to address complex challenges in quantum state tomography\cite{ma_how_2021,helsen_general_2022}, and supervised learning approaches have demonstrated impressive performance in entanglement classification, even for multi-class problems in high-dimensional systems \cite{vintskevich_classification_2023}. Experiments have demonstrated multipartite entanglement among up to 25 individually addressable atomic-ensemble interfaces\cite{pu_25interfaces_2018} and a random-access  quantum memory with 105 stored optical qubits\cite{jiang_105random_2019}, which highlight the growing complexity of engineered quantum systems. However, conventional entanglement-characterization methods, as well as most ML-based classifiers, require very large training datasets, thereby shifting the experimental burden to data acquisition. Although prior ML approaches have advanced entanglement classification, the critical challenge of achieving sample efficiency-reliable performance with minimal data-remains largely unaddressed. Therefore, developing learning models that align with the stringent data constraints of quantum experiments is imperative.\\

To address this, we propose a tailored hybrid neural network architecture specifically designed for sample-efficient entanglement classification: CNNs extract local, spatially invariant features from measurement data, while BiLSTMs capture complex sequential dependencies, together enabling robust pattern recognition from very few examples. We investigate two specific fusion paradigms to optimize this integration. When sufficient data is available, our model demonstrates exceptional performance, achieving accuracy exceeding 99.99$\%$ for classifying 3-qubit and 4-qubit entanglement. Most significantly, under extreme data scarcity—with merely 100 training samples—it maintains accuracy above 90$\%$, demonstrating rapid convergence and superior robustness compared to standard models. This work establishes a practical pathway toward scalable, data-efficient entanglement classification for complex quantum systems.

\section{Machine Learning Model}
\subsection{Basic Network Components}
Convolutional Neural Networks (CNNs), one of the most widely used deep neural network architectures, are designed to extract hierarchical features through successive convolution and pooling operations applied to input feature maps. These features are then mapped to fully connected layers to produce the final output. Both convolutional and fully connected layers typically employ activation functions (e.g., ReLU or LeakyReLU) to model nonlinear relationships. Unlike multi-layer perceptrons (MLPs), which rely entirely on fully connected layers, CNNs utilize local connectivity and weight sharing. This design significantly reduces the number of trainable parameters, enhancing model generalization and computational efficiency. Consequently, CNNs mitigate the overfitting problem.\\

\begin{figure}[htbp]
\centering
\includegraphics[width=1.0\columnwidth]{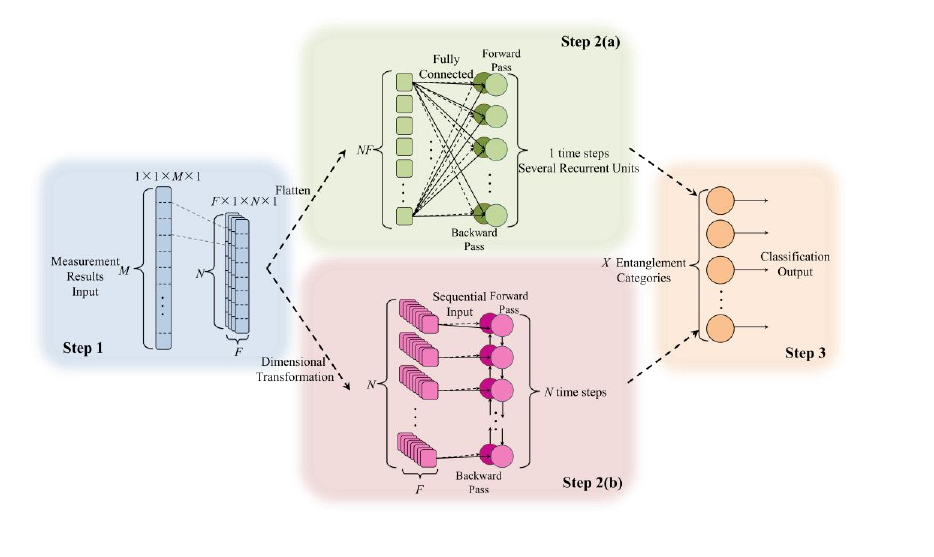}
\caption{The data transmission process of the two CNN-BiLSTM architectures. Step 1: Measurement outcomes of the quantum state are organized and fed into convolutional layers for initial feature extraction. Step 2: The convolutional features are processed in two ways. (a) \textbf{Architecture 1}: features are flattened and passed directly into BiLSTMs layers. (b) \textbf{Architecture 2}: a dimensionality transformation is applied: the feature maps after convolution and pooling (with a single timestep) are reshaped into a sequence, which is then processed by BiLSTMs layers in both forward and backward directions. Step 3: The processed data are integrated and mapped to a fully connected layer with softmax activation for final classification.}
\label{Fig1}
\end{figure}

Unlike MLPs and CNNs, where each input-output pair operates independently without explicit temporal dependencies, these architectures struggle with tasks involving strong sequential relationships. In contrast, Recurrent Neural Networks (RNNs) incorporate recurrent units that process current inputs while training states from previous time steps, making them particularly well-suited for sequential information processing. The RNNs' spatial unfolding structure with a single recurrent hidden layer is shown in  Figure\ref{Fig1}, where \textbf{X} and \textbf{Y} represent the input and output data vectors, respectively, while \textbf{H} denotes the output value vector of the recurrent unit. The weight matrix and bias vector are indicated by \textbf{W} and \textbf{B}, which can be wrapped in a non-linear activation function $f$. \\

At timestep $t$, the recurrent unit receives $\mathbf{W_{1}X^{t}}$ from the input layer, together with $\mathbf{W_{3}H^{t-1}}$ from its output at the previous timestep $t-1$:\\

\begin{equation}
\mathbf{H}^{(t)} = f_1\left([\mathbf{W}_1 \mathbf{X}^{(t)}, \mathbf{W}_3 \mathbf{H}^{(t-1)}] + \mathbf{B}_{1,3}\right)
= f_1\left(\mathbf{W}_{1,3} [\mathbf{X}^{(t)}, \mathbf{H}^{(t-1)}] + \mathbf{B}_{1,3}\right)
\end{equation}

where
\[
\mathbf{W}_{1,3} = \begin{bmatrix} \mathbf{W}_1^\top, \mathbf{W}_3^\top \end{bmatrix}^\top.
\]

Subsequently, $\mathbf{H}^{(t)}$ will be mapped to the output layer:

\begin{equation}
\mathbf{Y}^{(t)} = f_2\left(\mathbf{W}_2 \mathbf{H}^{(t)} + \mathbf{B}_2\right)
\end{equation}

Specifically, for the recurrent unit at the initial moment $t = 0$, we first define that $\mathbf{H}^{(t-1)} = \mathbf{0}$. Therefore:

\begin{equation}
\mathbf{H}^{(0)} = f_1\left(\mathbf{W}_1 \mathbf{X}^{(0)} + \mathbf{B}_{1,3}\right)
\end{equation}

At every time step following $t=0$, the output value $\mathbf{Y}^{(t)}$ is determined by the current input value $\mathbf{X}^{(t)}$ and is also influenced by the previous time step, which imparts RNNs a certain ``memory'' capability. This allows it to process sequential data effectively. Depending on the specific task requirements, the RNNs can either produce outputs at each moment in sequence or only provide the final result. Furthermore, by modifying the internal structure of the recurrent units, the ability of RNNs to capture the features of complex long sequences can be significantly enhanced. Common types include the Gated Recurrent Unit (GRU), Long Short-Term Memory (LSTM), Bidirectional Recurrent Neural Network (BiRNN), as well as their similar variants such as BiGRU and BiLSTM.

\subsection{Tailored Fusion Architecture for Sample-Efficient Learning}

Our proposed model is a concrete instantiation of the CNN-RNN paradigm, specifically designed for the sample-efficient classification of quantum entanglement. The overall pipeline is depicted in Figure~\ref{Fig1}. We implement this fusion through two main paradigms: Convolutional RNN (ConvRNN) architectures and CNN-RNN architectures. As shown in Equations (1)--(3), the main computational process within the fully connected RNN (FC-RNN) involves fully connected matrix multiplication. In ConvRNN, this multiplication is replaced by convolution operations, enabling effective extraction of multi-dimensional sequential features. Building on this, the RNN in ConvRNN can be adapted to create variations such as ConvGRU and their bidirectional variants.

Furthermore, these models can be categorized by their convolutional dimensionality. The second paradigm, CNN-RNN, adopts a sequential architecture that connects a CNNs feature extractor with a FC-RNN module, where the latter incorporates a key dimensionality transformation tailored for sample efficiency.

\subsubsection{1. Shared Backbone: Feature Extraction Modules}

The input is a one-dimensional vector composed of measurement outcomes. The CNNs module comprises 2--3 one-dimensional convolutional layers activated by the LeakyReLU function capturing local correlations across adjacent measurement bases for the first time. Then comes the second feature extraction by the BiLSTMs module. This configuration provides sufficient complexity to model complex sequential dependencies while maintaining a relatively critical generalization from few samples due to convolutional layers.

\subsubsection{2. Two Fusion Paradigms: Archi1 vs.\ Archi2}

The key innovation lies in how we connect the CNNs and BiLSTMs modules.

\begin{itemize}
    \item \textbf{Architecture 1 (Archi1): Feature-Flattening Fusion.} We utilize the $M$ measurement bases to measure an unknown quantum state. The measurement outcomes are organized as a vector of length $M$, which serves as the input for a single time step and a single feature map, resulting in input data dimensions of $1 \times 1 \times M \times 1$. In the CNNs component, performing multiple convolution and pooling operations, the dimensions are reduced to obtain $F$ one-dimensional feature maps, each containing $N$ features, resulting in the data with one single timestep and data dimensions of $F \times 1 \times N \times 1$. Then the output feature maps from the CNNs module will be directly flattened into a one-dimensional vector with $FN$ parameters. The resulted vector will then be fully connected linked into the BiLSTMs. This is a common and time efficient approach, but this inevitably ignores some of the before and after correlation the input of measurement results.

    \item \textbf{Architecture 2 (Archi2): Dimensionality-Transforming Fusion.} Considering that different feature maps reflect various characteristics extracted from the measurement data through down-sampling, and since the measurement bases are not completely independent, there are correlations between different measurement outcomes. To preserve the temporal sequential correlations among the measurement outcomes, we adopted a critical dimensionality transformation as opposed to flattening operations. The output feature maps with dimensions of $F \times 1 \times N \times 1$ from CNNs are then segmented, and the original feature dimension will be converted into a timestep dimension; the original feature map dimension will be transformed into a feature value dimension while keeping the total data amount unchanged. Ultimately, a sequence with $N$ timesteps is produced that each timestep contains $F$ feature values, with data dimensions of $N \times F$, and these data will be processed as a sequence by CNN-BiLSTM next.
\end{itemize}

\subsubsection{3. Output and Training Configuration}

The output from the BiLSTMs is passed through a final output layer with softmax activation for classification. The model is trained using the Adam optimizer with an initial learning rate of 0.01, and cross-entropy as loss function.

\section{Numerical Results}

\subsection{Dataset Generation}

Quantum states are categorically classified into only two distinct types in 2-qubit systems: separable states and entangled states. However, the classification of entanglement becomes significantly more complex for systems of larger multipartite systems. Crucially, two quantum states possess equivalent entanglement if they can be mutually transformed via local operations and classical communication (LOCC)\cite{bennett_mixed_1996} with non-zero probability. In 3-qubit and 4-qubit systems, quantum states fall into distinct equivalence families under Stochastic Local Operations and Classical Communication (SLOCC) \cite{dur_three_2000,verstraete_four_2002,pu_wstate-2025} equivalence classes. In 3-qubit systems, quantum states are classified into six equivalence families under SLOCC: separable states, three types of bipartite entangled states (A-BC, B-AC, C-AB), GHZ states, and W states. For the 4-qubit systems, ten types of SLOCC equivalence families are identified.

We evaluate the proposed CNN-BiLSTM architectures on two representative multipartite entanglement classification tasks: 3-qubit and 4-qubit systems. For each system, we generate a balanced dataset total of 400\,000 labeled quantum states for 3-qubit and 4-qubit systems, equally distributed among the relevant SLOCC equivalence classes (see Appendix~I for state parametrization). An independent set of 400\,000 samples is held out for testing. Each quantum state is encoded as a fixed-length vector of simulated projective measurements performed in a complete set of mutually unbiased bases.

\subsection{Performance with Abundant Training Data}

\begin{figure}[htbp]
\centering
\includegraphics[width=1.0\textwidth]{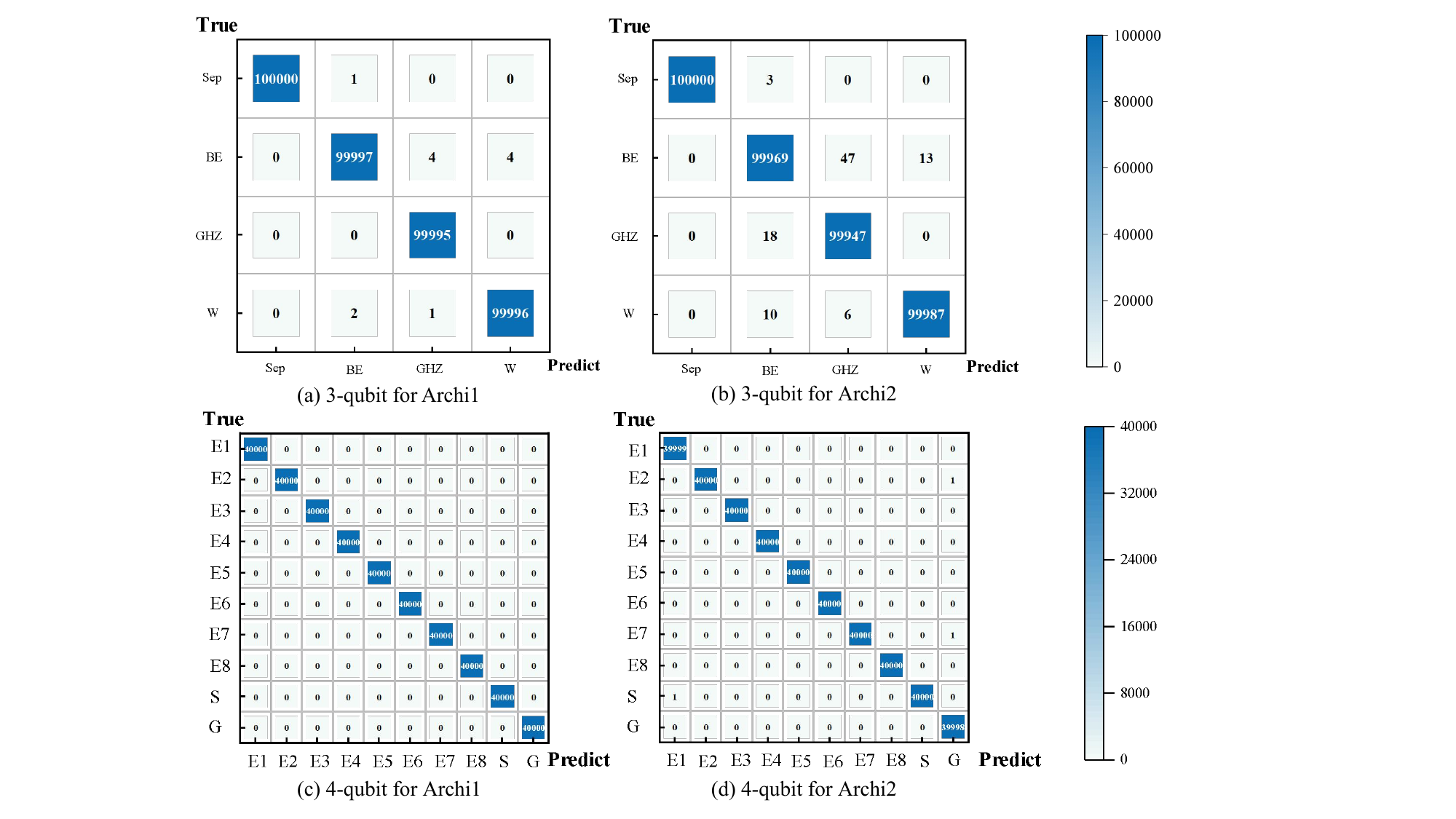}
\caption{Confusion matrices for 3-qubit and 4-qubit on two kinds of CNN-BiLSTM architectures. (a) 3-qubit for Archi1. (b) 3-qubit for Archi2. (c) 4-qubit for Archi1. (d) 4-qubit for Archi2.}
\label{fig:confusion}
\end{figure}

When trained on the full dataset (400,000 samples) for over 800 epochs, both fusion architectures achieve near perfect classification accuracy. As shown in the confusion matrices of Figure~\ref{fig:confusion}, Archi1 reaches 99.997\% (3-qubit) and 100.000\% (4-qubit), while Archi2 attains 99.976\% and 99.999\%, respectively. The near-perfect diagonals in the confusion matrices indicate that misclassifications are exceedingly rare and primarily occur between entanglement families with analogous local correlation patterns.

Notably, the training time of Archi2 (25\,h\,38\,m\,30\,s) exceeds that of Archi1 (2\,h\,33\,m\,39\,s) by approximately an order of magnitude. This reflects the additional computational overhead from the BiLSTMs' sequential processing of the reshaped feature sequence, attributable to the Back-propagation Through Time (BPTT) algorithm, which requires unrolling the network sequentially and computing gradients stepwise, thereby increasing computational complexity during the training and reducing training speed.

\subsection{Sample-Efficiency under Data Scarcity}

\begin{figure}[htbp]
\centering
\includegraphics[width=1.0\textwidth]{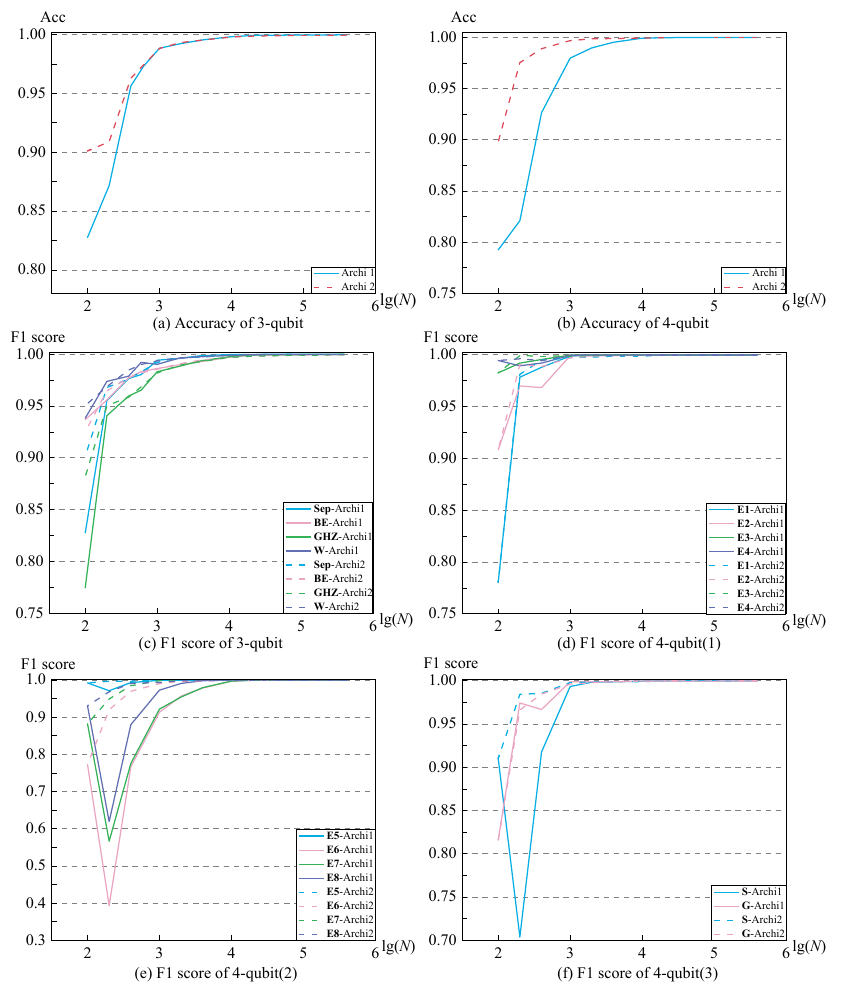}
\caption{The relationship between models' classification performance and training sample size (plotted against $\log N$). (a) accuracy of 3-qubit. (b) accuracy of 4-qubit. (c) F1 score of 3-qubit. (d-f) F1 score of 4-qubit part 1/2/3.}
\label{fig:relation}
\end{figure}

To assess sample efficiency, we systematically reduce the training set size while evaluating model performance on the fixed test set. As shown in Figure~\ref{fig:relation}(a--b), both architectures exhibit a monotonic decrease in accuracy with diminishing data, but Archi2 demonstrates a markedly slower decay. Crucially, with only 100 training samples, Archi2 maintains accuracy above 90\% for both system sizes, which indicates superior fitting capability and generalization performance in this data scarcity regime.

This superiority is further corroborated by the per-category F1 scores shown in Figure~\ref{fig:relation}(c--f). Under data scarcity, Archi2 consistently achieves higher F1 scores across all entanglement families, confirming its enhanced ability to balance precision and recall with limited training data. Additional metrics including true positive rate (TPR) and false positive rate (FPR) are provided in Supplement~II, which similarly favor Archi2 under data scarcity.

\subsection{Convergence Dynamics with Minimal Training Samples}

\begin{figure}[htbp]
\centering
\includegraphics[width=1.0\textwidth]{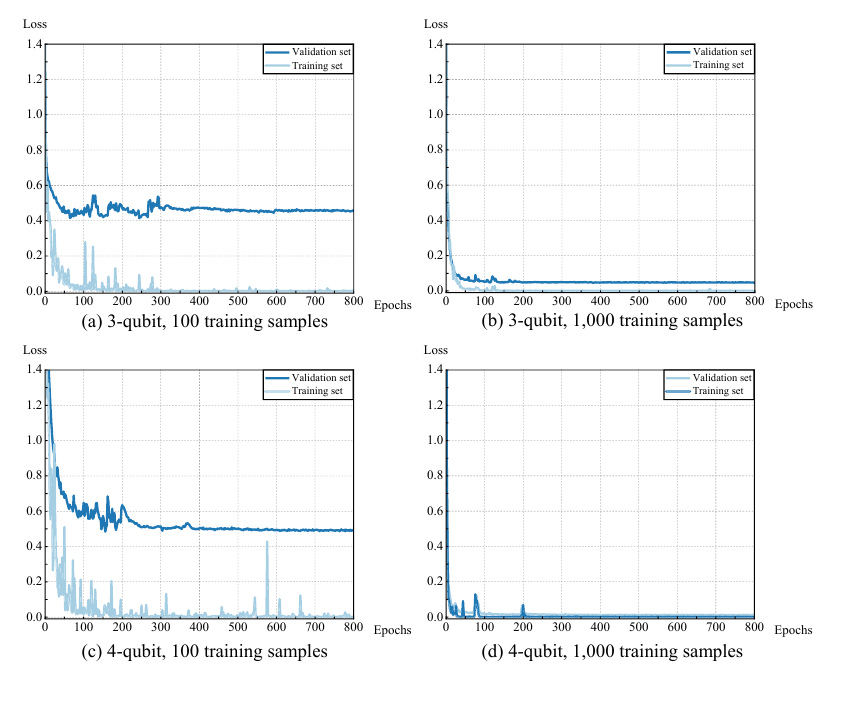}
\caption{The relationship between cross-entropy loss function value and training epochs. (a) 3-qubit case with 100 training samples. (b) 3-qubit case with 1\,000 training samples. (c) 4-qubit case with 100 training samples. (d) 4-qubit case with 1\,000 training samples.}
\label{fig:loss}
\end{figure}

We examine the training dynamics of Archi2 under extreme data scarcity (100 and 1,000 samples). Figure~\ref{fig:loss} plots the cross-entropy loss versus training epochs. With only 100 samples, the loss rapidly decays within the first 50 epochs and stabilizes near epoch~100, yielding final test accuracies of 90.33\% (3-qubit) and 91.00\% (4-qubit). When 1,000 samples are available, convergence accelerates substantially, reaching 98.79\% (3-qubit) and 99.74\% (4-qubit) accuracy by epoch~50. These results highlight the model's capacity to extract robust features and converge reliably even with very few training samples.

In terms of training time under scarcity, Archi2 requires 23\,s (3-qubit) and 98\,s (4-qubit) for 100 epochs with 100 samples, compared to Archi1's 16\,s and 38\,s, respectively. The moderate increase in computational time is outweighed by the substantial gain in accuracy, making Archi2 a practical choice for experimental settings where data acquisition is costly.

\subsection{Benchmarking Against Baseline Models}

\begin{figure}[htbp]
\centering
\includegraphics[width=1.0\textwidth]{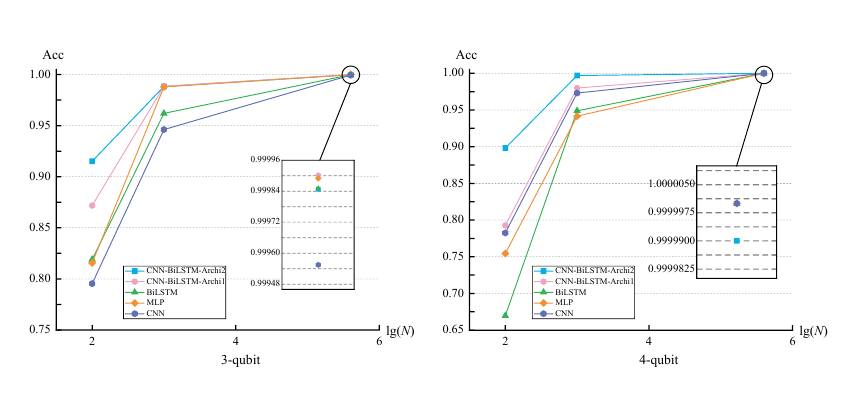}
\caption{Comparison of different models' accuracy as a function of training sample size.}
\label{fig:acc}
\end{figure}

We conduct a comprehensive comparison of our CNN-BiLSTM architectures against three standalone deep learning models: CNNs, BiLSTMs, and MLPs. Table~\ref{tab:time} summarizes the training times under both full-data and low-data conditions. Figure~\ref{fig:acc} compares the classification accuracy as a function of training sample size across all models.

\begin{table}[htbp]
\centering
\caption{Comparison of different models' training time.}
\label{tab:time}
\begin{tabular}{lcccc}
\toprule
Model & \multicolumn{2}{c}{400,000 samples} & \multicolumn{2}{c}{1,000 samples} \\
\cmidrule(lr){2-3} \cmidrule(lr){4-5}
& 3-qubit & 4-qubit & 3-qubit & 4-qubit \\
\midrule
Archi~2 & 2\,h\,38\,m\,45\,s & 25\,h\,38\,m\,30\,s & 37\,s & 42\,s \\
Archi~1 & 1\,h\,12\,m\,56\,s & 2\,h\,33\,m\,39\,s & 25\,s & 39\,s \\
BiLSTMs & 55\,m\,1\,s & 2\,h\,48\,m\,55\,s & 24\,s & 43\,s \\
MLPs & 10\,m\,10\,s & 33\,m\,37\,s & 20\,s & 19\,s \\
CNNs & 1\,h\,27\,m\,23\,s & 1\,h\,46\,m\,23\,s & 15\,s & 15\,s \\
\bottomrule
\end{tabular}
\end{table}

The comparative analysis reveals that with sufficient training samples, all evaluated models achieve comparable performance. Among them, however, the CNN-BiLSTM models (particularly Archi2) lead to substantially longer training time. In data-scarce scenarios, a pronounced difference emerges: the CNN-BiLSTM architecture demonstrates superior fitting capability overall. Notably, the CNN-BiLSTM-Archi2 variant achieves the highest accuracy with only a moderate increase in time overhead, showcasing exceptional overall performance under such constraints.

Although Archi2 achieves superior sample efficiency and classification accuracy under data-scarce conditions, it requires significantly longer training times than Archi1 and other baseline models. This overhead originates primarily from the Backpropagation Through Time (BPTT) algorithm used in BiLSTMs training. In Archi2, the dimensionality transformation expands the spatial length of the CNNs feature maps into the temporal dimension of the BiLSTMs, thereby increasing the effective sequence length. Because the computational and memory costs of BPTT scale with sequence length, this design choice inherently raises the training complexity.

Nevertheless, this trade-off is well justified within the paradigm of sample-efficient learning. Crucially, in experimental quantum tomography, the dominant cost typically lies in data acquisition, rather than computation. Preparing highly controlled quantum states, performing repeated measurements, and managing environmental noise are time-intensive and resource-limited processes. Reducing the required number of measurement samples by orders of magnitude—as demonstrated by Archi2—far outweighs the added computational overhead during training. Thus, the architecture makes a pragmatic trade-off: it shifts complexity from the experimental domain (data scarcity) to the computational domain (increased training time), where it is more manageable.

Several engineering strategies could be employed to mitigate the training time without compromising sample efficiency: adopting truncated BPTT to limit memory usage; leveraging hardware acceleration (e.g., GPU/TPU optimizations for long sequences); or exploring lightweight recurrent units (e.g., GRU or SRU) that preserve sequential modeling capacity with fewer parameters.Embedding such optimizations would facilitate the deployment of this sample-efficient framework in larger-scale quantum characterization experiments.

\subsection{Noise Robustness}

To assess the practical applicability of our CNN-BiLSTM architectures in realistic experimental environment, we evaluate their robustness against two prevalent noise types: dephasing noise and random measurement noise. In quantum experiments, dephasing noise arises from environmental interactions causing loss of phase coherence, while random noise stems from finite statistical sampling in projective measurements.

Dephasing noise primarily affects the phase of a quantum bit rather than its amplitude. We model dephasing noise using the phase damping channel with strength parameter~$\varepsilon$, where the off-diagonal elements of the density matrix decay as $\rho_{ij}\!\rightarrow\!(1\!-\!\varepsilon)\rho_{ij}$. For random noise, we simulate finite measurement statistics by sampling with $N_{\mathrm{measurement}}$ trials per measurement basis. For each situation, we use 40\,000 samples as testing set.

\begin{figure}[htbp]
\centering
\includegraphics[width=1.0\linewidth]{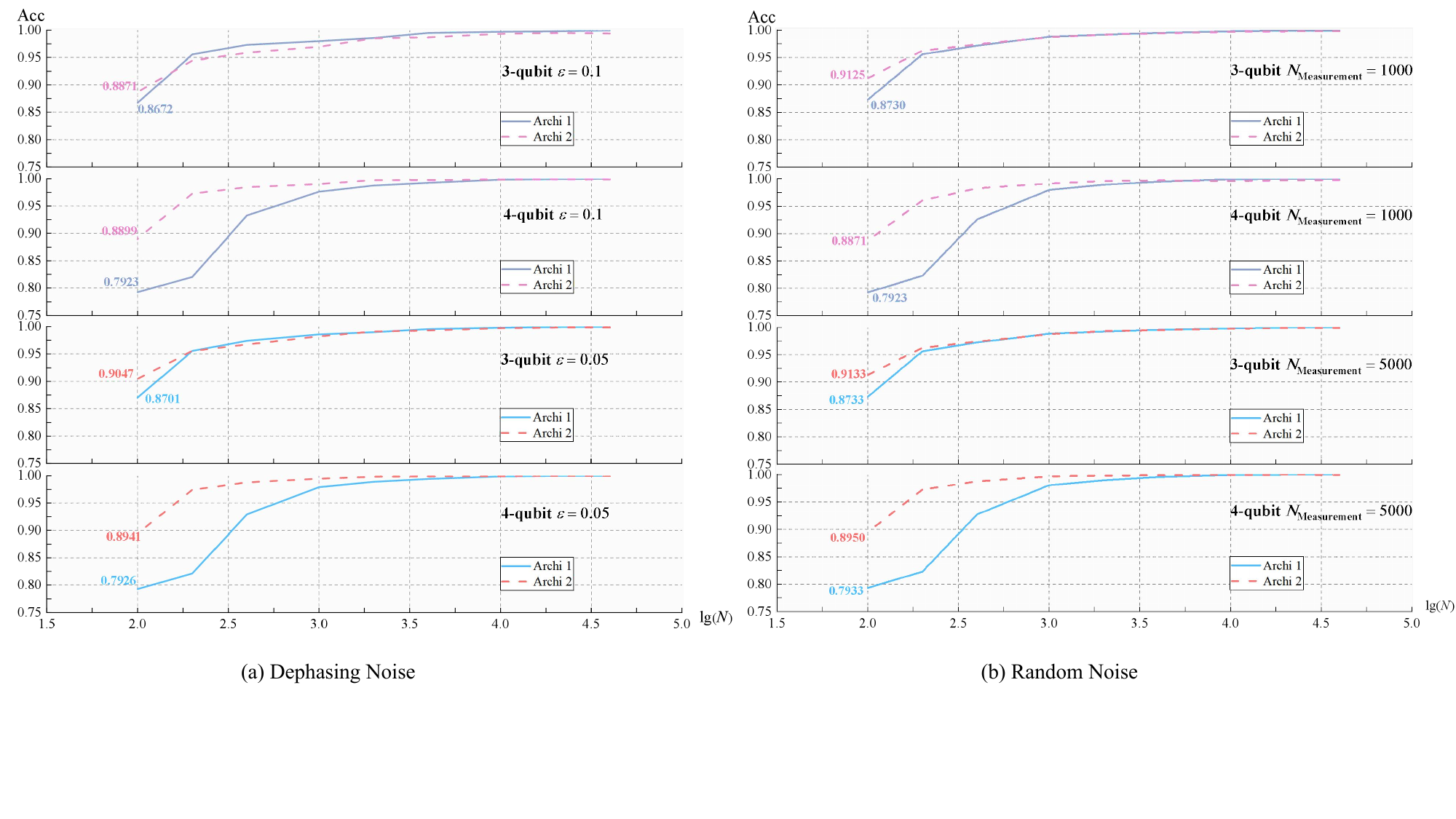}
\caption{Robustness of CNN-BiLSTM architectures to experimental noise, training sample size N is plotted as $\log N$. (a) Classification accuracy under dephasing noise of strength~$\varepsilon=0.05, 0.1$. (b) Classification accuracy under random statistical noise obtained from different measurements.}
\label{fig:noise}
\end{figure}

Figure~\ref{fig:noise} presents the classification accuracy of both architectures under dephasing noise and random measurement noise. Remarkably, even with only 100 training sample size, Archi2 maintains accuracy above 88\% whereas Archi1 drops below 80\% under identical conditions. As training sample size N increases from 100 to 40\,000, both architectures converge toward their noiseless performance. This superior noise resilience stems from Archi2's dimensionality transformation, which preserves correlated features across measurement bases and effectively filters out incoherent noise contributions. And this confirms that the sequential feature processing in Archi2 provides inherent denoising capabilities by exploiting temporal correlations in measurement data.

These results significantly broaden the applicability of our sample-efficient framework. The demonstrated tolerance to realistic noise levels—without requiring noise-aware training or additional regularization—validates that the four-orders-of-magnitude reduction in training data demand does not compromise experimental feasibility. Rather, the architecture's intrinsic structure provides robustness that compensates for limited training data and noisy inputs, establishing a practical pathway for deployment in current noisy intermediate-scale quantum (NISQ) devices.

\section{Discussion}

The motivation for the “dimensionality transformation” in Archi2 is grounded in the physics of quantum measurement and the requirement for data-efficient learning. Physically, the outcomes from complementary measurement bases are inherently interdependent due to quantum uncertainty and the non-commutativity of measurement operators. The CNNs' output feature maps are thus not independent scalars, but a set of correlated vectors that encode the state's local entanglement patterns from distinct filter-induced “perspectives”. The conventional flattening operation severs these physically meaningful correlations, discarding vital quantum correlation information. 

Archi2 reconfigures these feature maps by treating the number of maps as the feature dimension and their spatial length as a sequence of timesteps. This structured reshaping produces an ordered sequence, enabling the subsequent BiLSTMs to explicitly model the contextual relationships between different feature patterns bidirectionally, thereby synthesizing a global, high-level representation of the quantum state. Theoretically, this physics-aware representation preserves inter feature correlations, yielding higher information density and lower redundancy than a flattened vector. Consequently, the model can extract more essential, physical patterns from each training sample. Under extreme data scarcity, this efficient information utilization directly enhances representational capacity and generalization, which is the foundation of superior sample efficiency we demonstrate.

\section{Conclusion}

Our machine learning model integrates powerful fitting capabilities for high classification accuracy with enhanced generalization for broad applicability in multi-qubit entanglement classification. The weight-sharing architecture of the CNNs bolsters the model's generalization performance, while the sophisticated recurrent units of the BiLSTMs excel at capturing complex data patterns. By synergistically integrating CNNs and BiLSTM, and specifically tailoring their interconnection based on the requirements of the entanglement detection problem, we have developed a powerful and versatile classification model for 3- and 4-qubit systems.

This study achieves a groundbreaking reduction in the required training samples by four orders of magnitude (from 400\,000 to merely 100), while maintaining high accuracy exceeding 90\%. The developed enhanced CNN-BiLSTM classification model delivers high-precision classification for 3- to 4-qubit pure quantum systems, even under the constraint of extremely limited training data. Notably, the rapid decline of the loss function value with increasing training epochs enable reliable classification results with a minimal number of epochs. This substantial reduction in training sample demand significantly alleviates the challenges and resource burden associated with data acquisition and processing, especially within high-dimensional quantum systems. Furthermore, the diminished requirement for extensive training epochs not only effectively shortens training time but also greatly enhances the flexibility in parameter adjustment during model training. Given the established correspondence between our defined measurement operators and the optical polarization states, this model is readily adaptable to practical experimental implementation, facilitating its deployment in applications.

While demonstrated on pure states of 3- and 4-qubit systems, our architecture is inherently scalable and adaptable. The CNN-BiLSTM fusion is not limited to specific system sizes or entanglement families. Future work may explore its application to mixed-state entanglement classification, noisy experimental data, or larger systems by incorporating attention mechanisms or graph neural networks to capture more complex quantum correlations. Furthermore, the principle of physics-aware representation reshaping---preserving measurement correlations in model-native structures---could inspire sample-efficient designs for other quantum learning tasks, such as quantum control optimization or variational quantum algorithm benchmarking.

{\small

\section{Acknowledgements}

\paragraph{Author Contribution Statement.}
Q.S constructed the model. YD.S provided the physical interpretation. Q.S,and YD.S analyzed the data. RQ.H handled the signal processing tasks and provided guidance on artificial intelligence-related aspects. Q.S, YD.S, Y.H, YH.M and N.J wrote the manuscript. N.J proposed and supervised the project. All authors read and approved the final manuscript.
\paragraph{Funding.} 
This work was supported by the National Natural Science Foundation of China (92476116, 12204055, 52504011); Fundamental Research Funds for the Central Universities (2243100012); CNPC Innovation Found(2025DQ02-0508).
\paragraph{Disclosures.} 
The authors declare no conflicts of interest.
\paragraph{Data Availability.} 
The datasets generated and analysed during the current study are not publicly available but are available from the corresponding author on reasonable request.
\paragraph{Code Availability.}
The underlying code for this study and training/validation datasets is not publicly available but may be made available to qualified researchers on reasonable request from the corresponding author. 
\paragraph{Supplemental Document.}
 See Supplementary for supporting content.
}

\printbibliography


\end{document}